\documentclass[a4paper]{jpconf}
\usepackage{graphicx}

\usepackage[colorlinks=true,linkcolor=blue]{hyperref}

\begin{document}
\title{A highly granular calorimeter concept for long baseline near detectors}

\author{Lorenz Emberger and Frank Simon\footnote[1]{Speaker}}

\address{Max-Planck-Institut f\"ur Physik, Munich, Germany}

\ead{fsimon@mpp.mpg.de}

\begin{abstract}
Future long baseline neutrino experiments such as the DUNE experiment under construction at Fermilab will perform precision measurements of neutrino oscillations, including the potential for the discovery of CP violation in the lepton sector. These measurements require an understanding of the unoscillated neutrino beam with unprecedented accuracy. This will be provided by complex near detectors which consist of different subsystems including tracking elements and electromagnetic calorimetry. A high granularity in the calorimeter, provided by scintillator tiles with SiPM readout as used in the CALICE analog hadron calorimeter, provides the capability for direction reconstruction of photon showers, which can be used to determine the decay positions of neutral pions. This can enable the association of neutral pions to neutrino interactions in the tracker volume, improving the event reconstruction of the near detector. Beyond photon and electron reconstruction, the calorimeter also provides sensitivity to neutrons. In this presentation, we will discuss a simulation study exploring the potential of high granularity for the electromagnetic calorimeter of the DUNE near detector. Particular emphasis will be placed on the combination with a high pressure TPC as tracking detector, which puts particularly stringent requirements on the calorimeter. The dependence of the projected detector performance on granularity, absorber material and absorber thickness as well as geometric arrangement satisfying the constraints of the TPC are explored.  
\end{abstract}

\section{Introduction}

LBNF/DUNE, the Deep Under Neutrino Experiment \cite{Acciarri:2016ooe} at the Long Baseline Neutrino Facility, currently under construction in the US, will pursue a broad program in neutrino physics and astrophysics. This includes the search for CP violation in the lepton sector and the determination of the mass ordering of the neutrinos by measuring the oscillations of accelerator-produced $\nu_\mu$ and $\bar{\nu}_\mu$ with a baseline of 1300\ km. The oscillation program relies on a far detector ultimately consisting of four \mbox{10\ kt} liquid argon TPCs deep underground in the Sanford Underground Research Facility at a distance of 1300\ km, and a near detector located on the Fermilab site 574\ m away from the production target. The near detector will consist of a liquid argon TPC and a down-stream magnetized multi-purpose detector system with a tracking detector also serving as neutrino target, surrounded by electromagnetic calorimetry and muon detectors. The design of the near detector is currently under study \cite{WeberND:2018}, with a straw tube tracker and a high pressure gaseous argon TPC under investigation for the tracking detector, possibly supplemented by a three-dimensional scintillator tracker. The option with a high pressure TPC puts more stringent requirements on the capabilities of the calorimeter, since the relatively low density of the gas results in a low conversion probability of photons, and thus makes the reconstruction of neutral pions with at least one converted photon detected in the tracker very inefficient. In addition, also the probability for a sizable signal from neutrons produced in the interaction is very low. This makes a calorimeter which is capable of determining the directions of photons and the vertex position of neutral pions that can thus be associated to neutrino interactions, as well as a detector with high efficiency for neutrons, interesting. A highly granular plastic scintillator-based sampling calorimeter may provide such capabilities. In contrast to the typical environment at collider detectors, which is characterized by highly energetic particles, the relevant particle energies here are rather low, with many neutral pions produced almost at rest or with kinetic energies of 300 to 400\ MeV for pions originating from resonance decays. This puts a premium on the calorimeter performance for photons in the energy range from 50 MeV to several hundred MeV. Still, given that the neutrino energies in DUNE extend to several GeV, the calorimeter performance up into the multi-GeV region remains relevant.

\section{A first concept of the calorimeter}

\begin{figure}
\centering
\includegraphics[width = 0.3\textwidth]{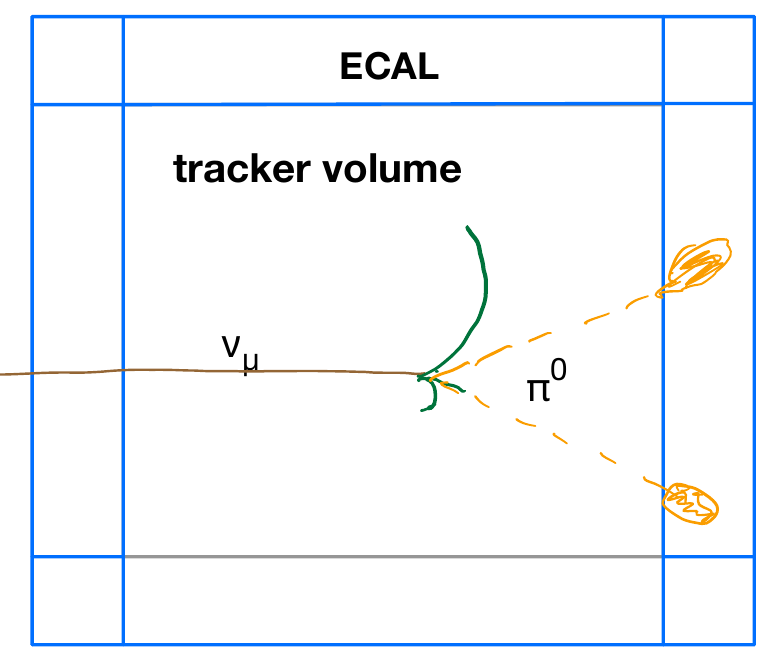} \hspace{10mm}
\includegraphics[width = 0.4\textwidth]{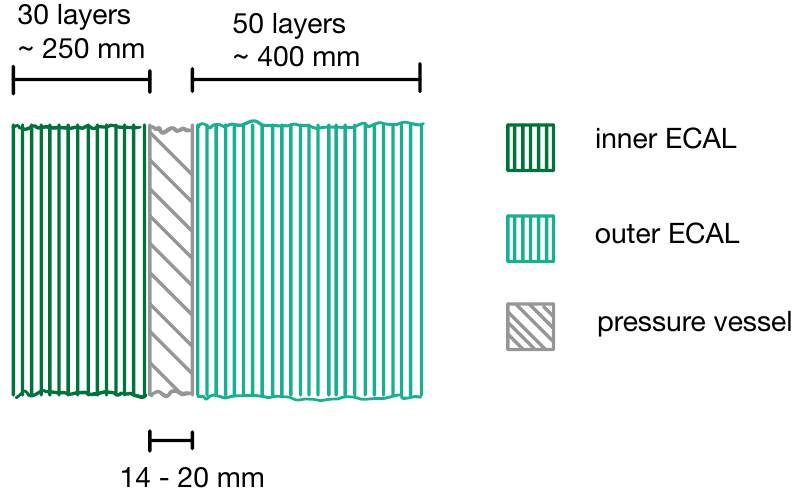}
\caption{\textit{Left:} Overall layout if the calorimeter studied in simulations, showing a cube-like structure with an edge length of the inner volume of 3 m. One of the goals of the detector, the association of neutral pions to a position within the tracker volume, is illustrated. \textit{Right:} Longitudinal structure of the calorimeter. Depending on precise layout and space constraints, the pressure vessel of the TPC may split the calorimeter into an inner and an outer segment. \label{fig:CaloLayout}}
\end{figure}

To investigate the potential of a highly granular calorimeter in the context of the DUNE near detector with a high pressure TPC, a first ``naive'' calorimeter layout has been studied in GEANT4 \cite{Agostinelli:2002hh}  based simulations, using the QGSP\_BERT\_HP physics list when simulating neutrons. The general layout of this concept is illustrated \autoref{fig:CaloLayout} \textit{left}. The calorimeter fully encloses the tracker volume, here assumed to be a cube with an edge length of 3 m. It should be noted that a realistic high-pressure TPC will have a somewhat larger volume, resulting in a larger calorimeter system than the one simulated here. 

A high pressure TPC, operated at a pressure on the order of 10\ bar, requires a robust pressure vessel to provide the necessary mechanical stability. While the location of the calorimeter outside of the pressure vessel considerably simplifies the integration of the system, the additional material between tracker and calorimeter introduced by the vessel, which corresponds to 0.4 to 1 $X_0$ depending on material and design, significantly degrades the performance of the calorimeter in particular for lower-energy photons. It is thus preferable to locate the entire calorimeter inside of the pressure vessel to avoid dead material between the tracking volume and the calorimeter. Depending on available space and integration considerations, this may be reduced to a partial inclusion of the ECAL, resulting in a split of the detector into an inner and an outer segment, as shown in  \autoref{fig:CaloLayout} \textit{right}.

\begin{figure}
\centering
\includegraphics[width = 0.7\textwidth]{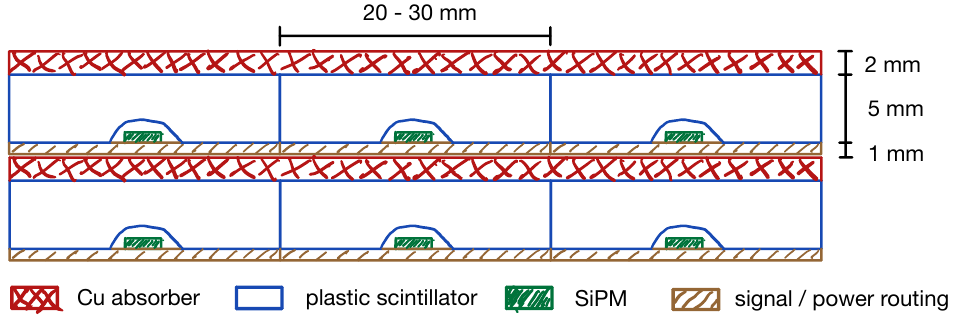} 
\caption{Illustration of the calorimeter layers, showing individual scintillator tiles with directly coupled silicon photomultipliers, absorber layers and additional space for signal and power routing. Here, the default scenario studied in this paper with 2 mm thick copper absorbers is shown, but other absorber thicknesses and materials have been studied as well. \label{fig:CaloLayer}}
\end{figure}

The calorimeter itself is a sampling calorimeter with alternating absorber and plastic scintillator layers. In addition, a gap to account for signal and power routing for the photon sensor is included in each layer, at present simulated by an air gap. The scintillator layer consists of individual scintillator tiles with a typical size of $20\times 20$ -- $40\times 40$\ mm$^2$, each read out by a directly coupled silicon photomultiplier, following designs developed in the context of the CALICE analog hadron calorimeter \cite{collaboration:2010hb, Simon:2010hf, Liu:2015cpe}. This technology is capable of providing nanosecond-level timing for individual detector cells \cite{Simon:2013zya}, providing the potential for precise time determination of photon clusters which can be used to assist the event reconstruction. The general layer structure of the detector is illustrated in \autoref{fig:CaloLayer}. In the present study, lead and copper with a thickness of 1\ mm and 2\ mm are investigated as absorber material, while the thickness of the plastic scintillator is always kept at 5 mm. The default scenario adopted here is 2\ mm copper absorbers. A total of 80 detector layers is assumed, resulting in a thickness of approximately 12 $X_0$.  

\section{Simulation studies}

The performance of the calorimeter concept was studied in simulations for single photons, neutral pions and neutrons \cite{ThesisEmberger}. To study the scaling of the detector performance with absorber material and thickness, granularity in different regions of the detector and the presence of a TPC pressure vessel, these parameters are varied. For the absorber, lead and copper with a thickness of 1\ mm and 2\ mm are studied. The granularity of the readout elements is varied from $10 \times 10 \ \mathrm{mm}^2$ to  $40 \times 40 \ \mathrm{mm}^2$, also considering different granularities in different depths of the detector. To study the impact of the pressure vessel, the performance without a vessel is compared to the one with the two different vessel types introduced above. These studies are performed for single photons only, for neutral pions and neutrons only the default detector design with 2 mm copper absorbers, $20 \times 20 \ \mathrm{mm}^2$ granularity and without pressure vessel is considered.

In the simulation, the energy deposited in each scintillator cell is stored, taking into account scintillator saturation effects. A very basic digitization is implemented, with a Poissonian smearing accounting for photon statistics assuming 25 detected photons per MeV, and a Gaussian noise of 100\ keV applied on each cell. A noise cut of 500\ keV is imposed on each cell following the application of smearing and noise before further analysis.

For the study of the pointing capability of the calorimeter, the reconstruction of the photon direction is needed. This is done in a two-stage process, with a principal component analysis of all hits in the detector to provide a first estimate of the main shower axis, followed by a straight-line fit through the center of gravities reconstructed in each layer. At present, a clustering algorithm that identifies detector cells belonging to one photon is not yet implemented. Instead, the truth information from the simulation is used to assign detector hits to photons when simulating neutral pions. 

In the following, selected results of the studies are presented and discussed.

\subsection{Single photons}

\begin{figure}
\centering
\includegraphics[width = 0.49\textwidth]{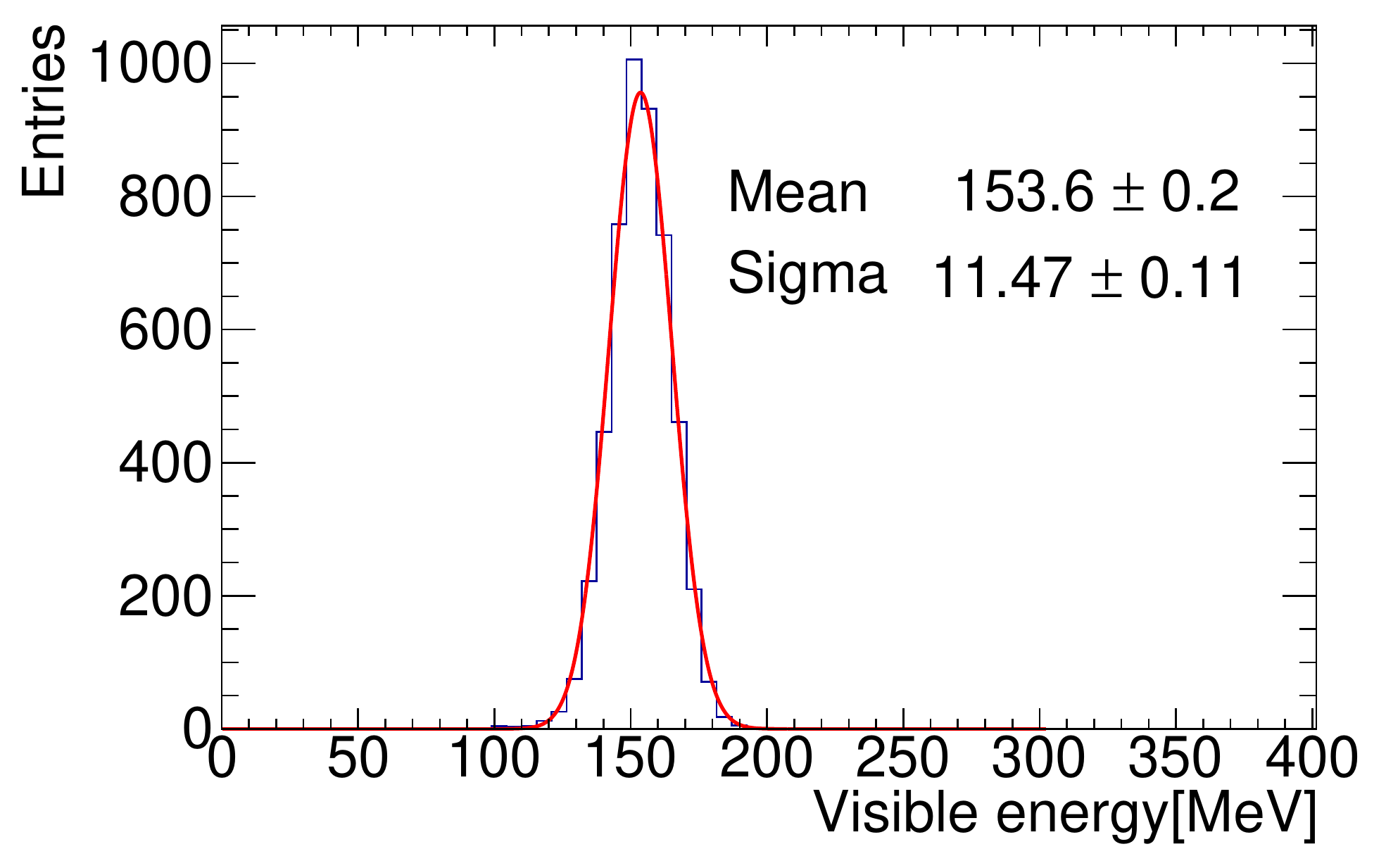} \hfill
\includegraphics[width = 0.49\textwidth]{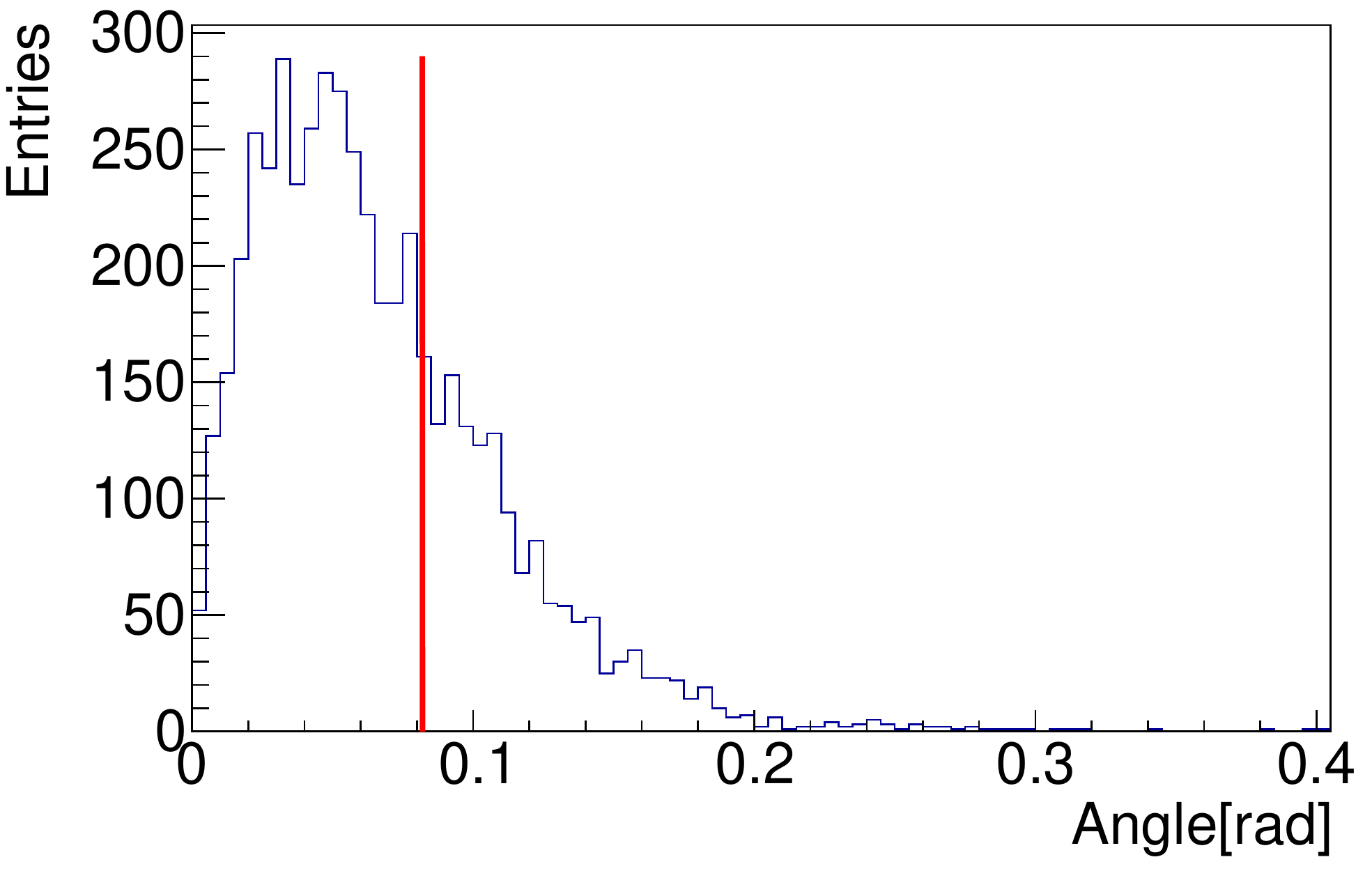}
\caption{Distribution of visible energy (left) and opening angle between reconstructed and true photon direction (right) for 450\ MeV photons in the default detector configuration. The energy resolution is given by sigma divided by the mean of the visible energy, while the angular resolution is given by the 68\%ile of the opening anlge between reconstructed and true photon direction, shown by the red vertical line in the right panel. \label{fig:SinglePhotonResolution}}
\end{figure}

For single photons the energy resolution and the angular resolution are taken as performance parameters. The energy resolution is defined as the sigma divided by the mean of a gaussian fit to the distribution of visible energy after applying the smearing steps and the cuts discussed above. The angular resolution is given by the 68\%ile of the opening angle between the reconstructed and the true photon direction. \autoref{fig:SinglePhotonResolution} shows the energy and angular resolution for 450\ MeV photons for the default detector configuration.

The energy dependence of both the energy and the angular resolution are well described by 
\begin{equation}
\frac{\sigma(E)}{E} = \frac{A}{\sqrt{E / \mathrm{GeV}}} \oplus \frac{B}{E / \mathrm{GeV}} \oplus C \ \mathrm{and} \  \Delta_{68\%} \Theta \mathrm{[rad]} = \frac{A}{\sqrt{E / \mathrm{GeV}} }\oplus \frac{B}{E / \mathrm{GeV}} \oplus C, \label{eq:Resolution}
\end{equation}
respectively. For the default detector configuration, the energy resolution follows \autoref{eq:Resolution} with $A = 5.2\%$, $C = 2.9\%$ and $B$ consistent with zero, while the angular resolution is described by $A = 0.045$, $B = 0.019$ and a negligible constant term $C$. The inclusion of additional material and material non-uniformities, as well as more realistic modeling of the detector response as discussed in \autoref{sec:realism} will result in a deterioration of these parameters.

\begin{figure}
\centering
\includegraphics[width = 0.49\textwidth]{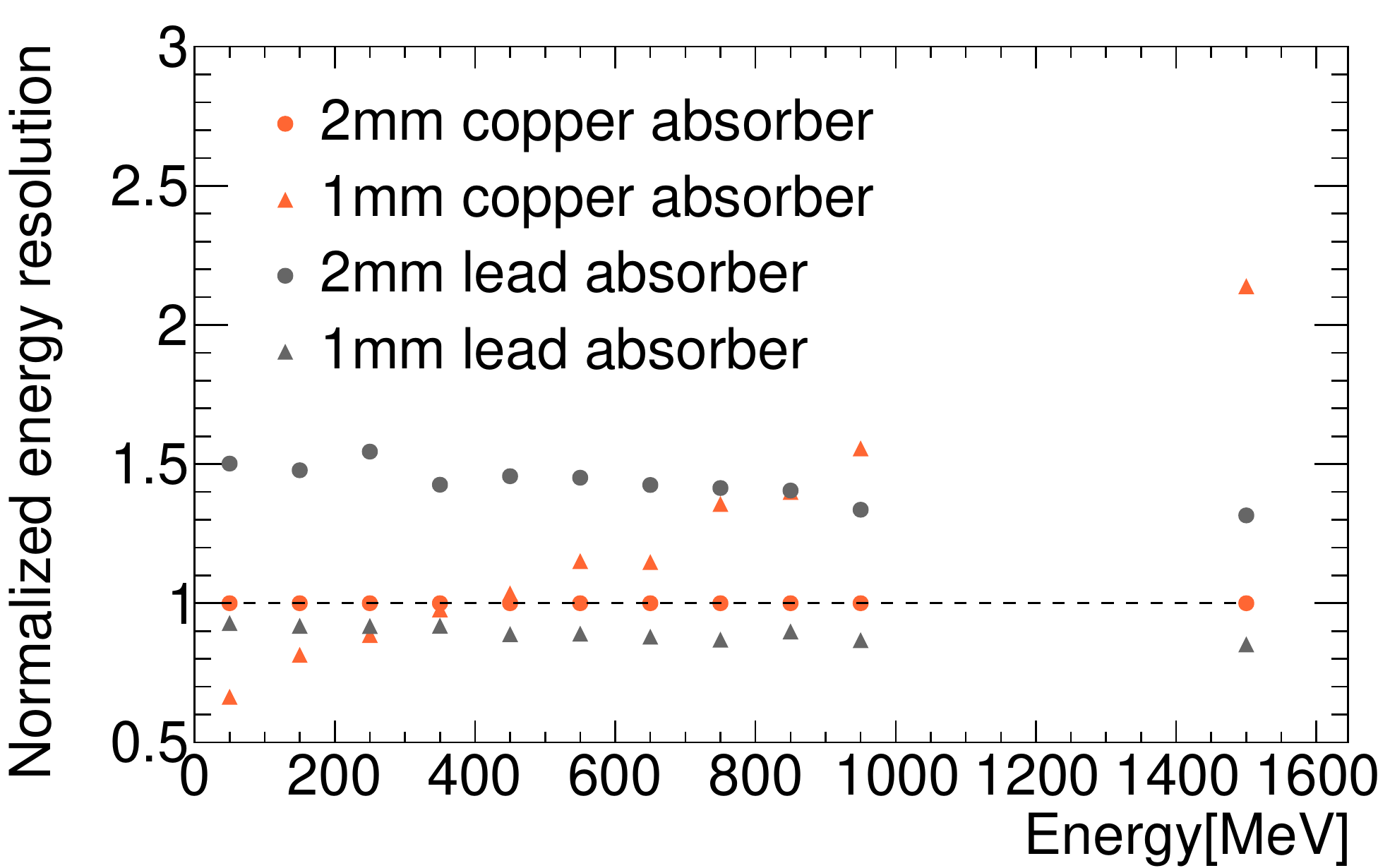} \hfill
\includegraphics[width = 0.49\textwidth]{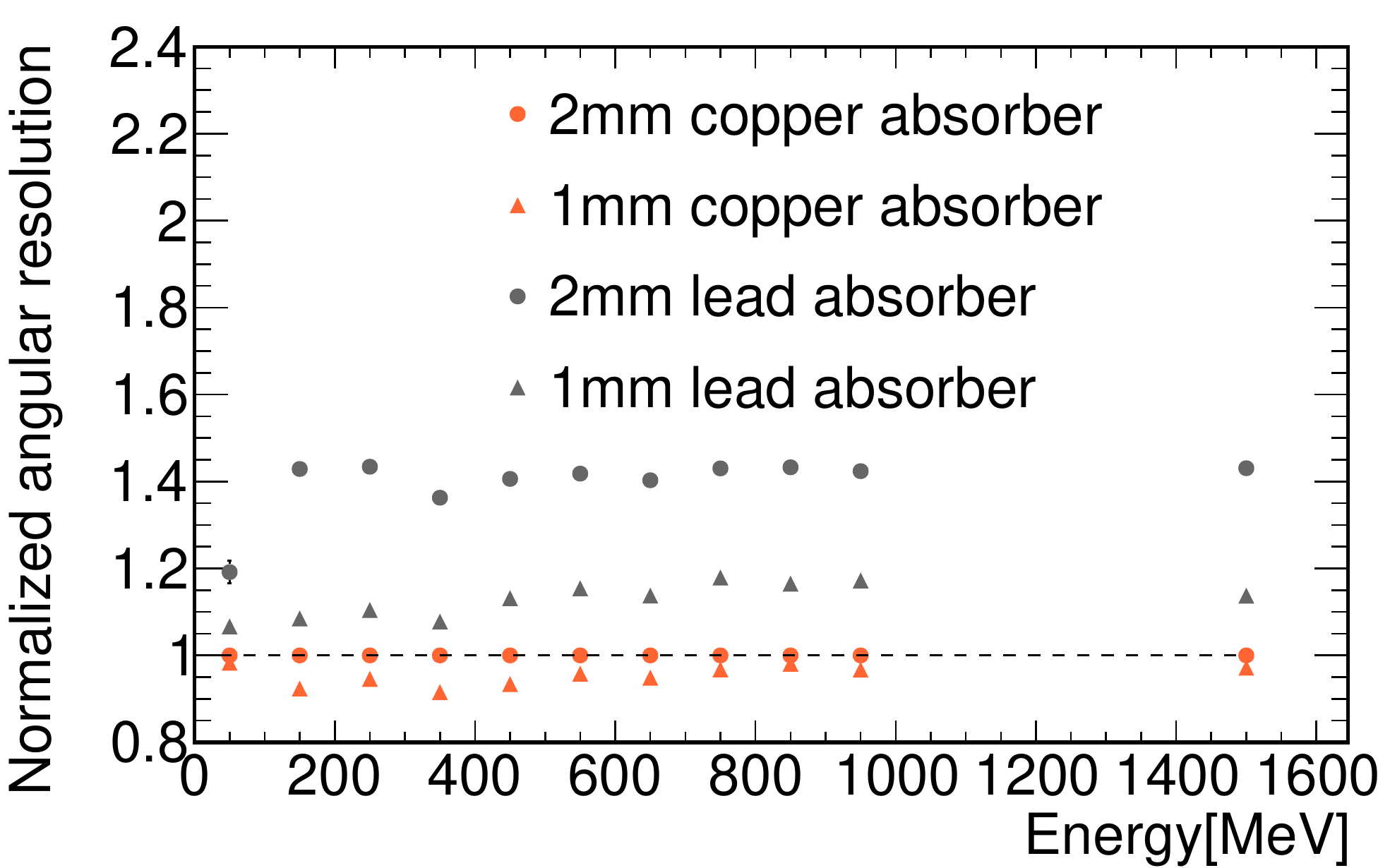}
\caption{Energy (left) and angular (right) resolution for different choices of absorber material and thickness, relative to the resolutions obtained for 2 mm thick copper plates. \label{fig:AbsorberInfluence}}
\end{figure}

The choice of the absorber material and thickness has a strong impact on the energy and angular resolution. In general, a higher sampling fraction, achieved by thinner absorber plates or by using a material with lower density / larger $X_0$, results in performance improvements. In the case of the energy resolution, these improvements are limited by longitudinal leakage, which deteriorates the performance for very thin absorbers. \autoref{fig:AbsorberInfluence} shows the influence of the choice of the absorber on energy and angular resolution. It is apparent that 1 mm thick copper absorbers are insufficient to contain showers even at rather moderate energies, resulting in poor energy resolution over a large part of the considered range, while 1 mm thick lead absorbers provide the best performance. Copper absorbers provide better angular resolution, which is due to the reduced Moliere radius and longer showers in copper compared to lead. Going below 2 mm thickness for copper yields only marginal improvement. 

\begin{figure}
\centering
\includegraphics[width = 0.49\textwidth]{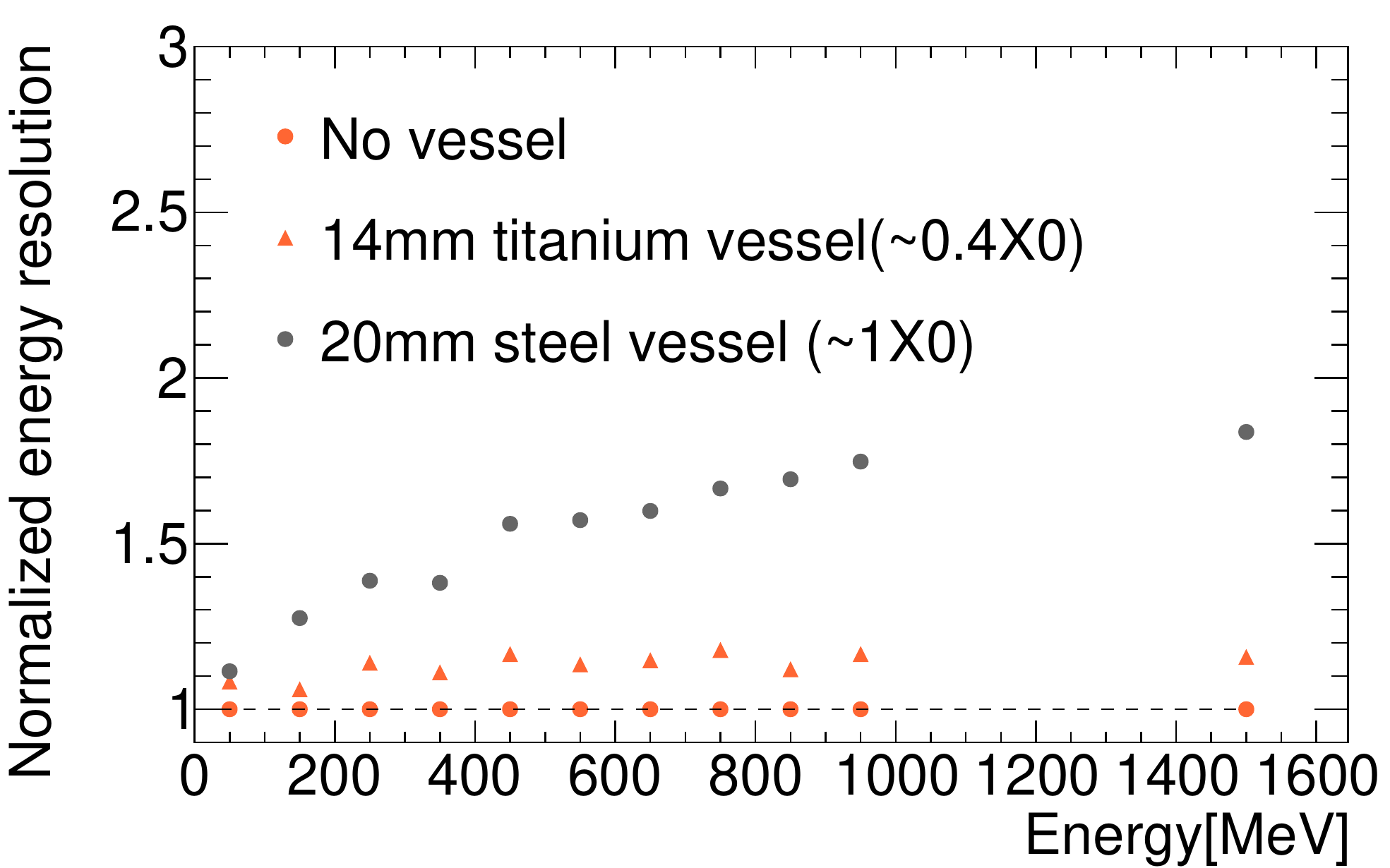} \hfill
\includegraphics[width = 0.49\textwidth]{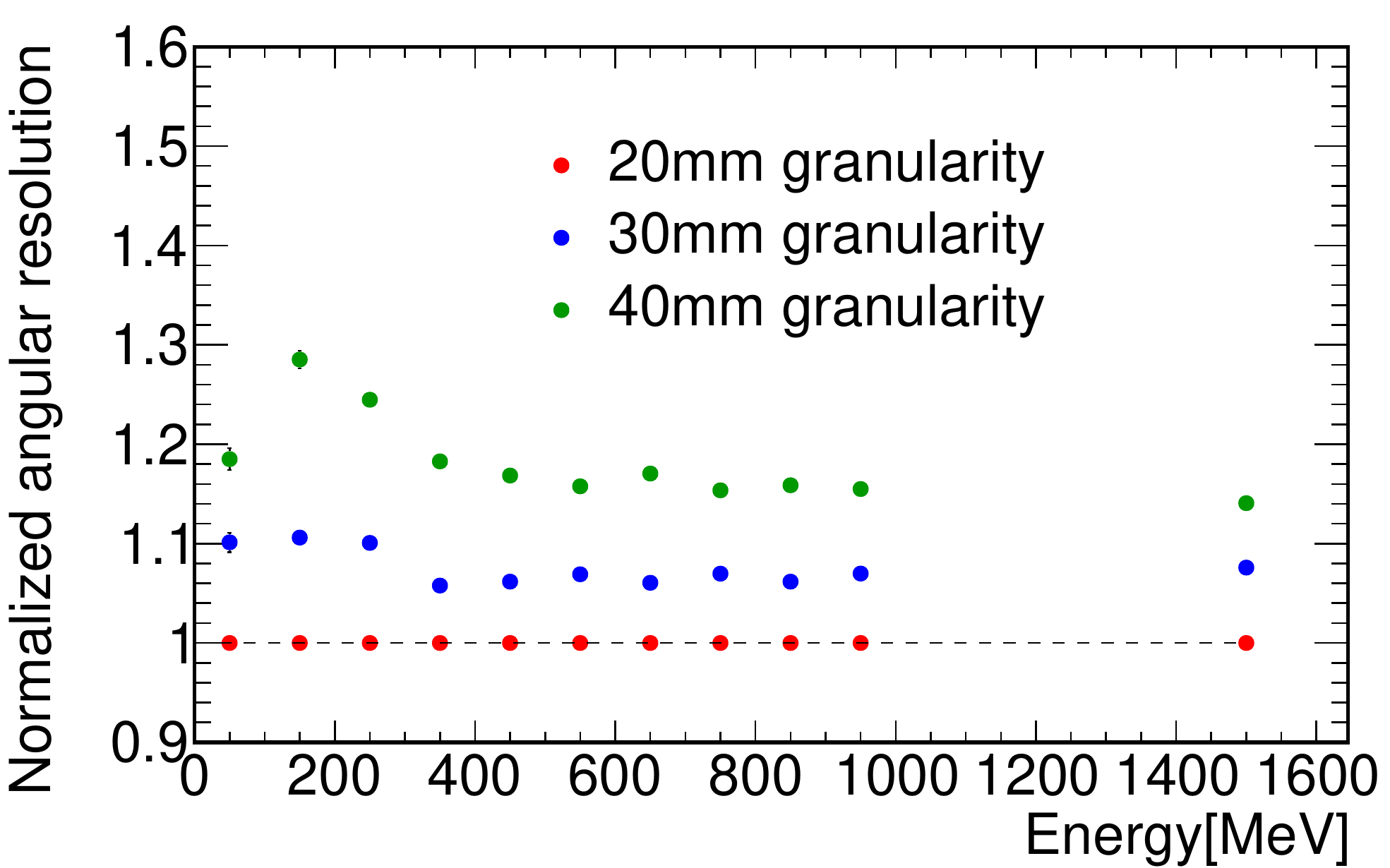}
\caption{Left: Change of energy resolution with the presence of a pressure vessel after the first 30 layers of the calorimeter, relative to the resolution without the vessel. The impact of the extra material on the angular resolution is negligible. Right: Angular resolution for different readout granularities, relative to the resolution obtained with a cell size of $20 \times 20 \ \mathrm{mm}^2$ . \label{fig:VesselandGranularity}}
\end{figure}

The presence of a pressure vessel (or some other structural passive material of comparable thickness) after the first 30 layers of the calorimeter has a significant impact on the energy resolution, in particular in the case of 20\ mm of stainless steel, as shown in \autoref{fig:VesselandGranularity} left. The negative impact of the presence of the extra material is considerably reduced in the case of 14\ mm of titanium. It should be noted that for simplicity a homogeneous thickness of the pressure vessel is assumed here. From an engineering point of view this is not required, and may also not be the most practical solution. A thinner shell with additional support ribs would result in less overall deterioration of the calorimeter performance, with large local disturbances in the region of the support structures. The impact of additional material after layer 30 on the angular resolution is negligible.  

The angular resolution strongly depends on the granularity of the readout, in particular in the first layers of the calorimeter. \autoref{fig:VesselandGranularity} right shows the change in angular resolution for coarser granularities relative to a cell size of $20 \times 20 \ \mathrm{mm}^2$ throughout the detector, demonstrating the benefit of higher granularity. Going below a cell size of $20 \times 20 \ \mathrm{mm}^2$ yields only a small additional benefit. The angular resolution profits in particular from a higher granularity in the first 10 layers of the detector, while the impact of the granularity beyond layer 30 is marginal.

\subsection{Neutral pions}

For the identification and localization of neutral pions both the energy and the angular resolution are relevant. Here, the capability of the default detector scenario with a granularity of $20 \times 20 \ \mathrm{mm}^2$, 2 mm thick copper absorbers and without a pressure vessel within the calorimeter volume is studied. For simplicity, only neutral pion decays into two photons, which make up approximately 99\% of all decays, are considered. The position of the neutral pion decay is determined by a minimization procedure which takes both the reconstructed energy and the reconstructed direction of the two photons into account. As a starting point, the position given by the point of closest approach of the two reconstructed photon directions projecting back into the detector volume enclosed by the calorimeter is taken. This position is then varied, minimizing the following equation:
\begin{equation}
\chi^2 = \frac{(m_{\mathrm{reco}} - m_{\pi^0})^2}{\sigma^2_{\mathrm{mass}}} + a \left[ \frac{\Delta\phi_1^2}{\sigma^2_{\mathrm{angle}_1}} + \frac{\Delta\phi_2^2}{\sigma^2_{\mathrm{angle}_2}} \right].
\end{equation}
Here, $m_{\mathrm{reco}}$ is the reconstructed mass of the neutral pion candidate taking the reconstructed energies of the two photons, as well as the opening angle given by the assumed particle decay position and the reconstructed positions of the center of gravity of the two photon clusters in the calorimeter. The expected mass resolution $\sigma_{\mathrm{mass}}$ is calculated from the energy resolution of the two photons. $\Delta\phi_1$ and $\Delta\phi_2$ are the angles between the direction from the assumed pion decay position to the respective center of gravity of the corresponding photon cluster and the reconstructed photon direction, while $\sigma_{\mathrm{angle}_1}$ and $\sigma_{\mathrm{angle}_2}$ are the respective angular resolutions determined from the energy of the photons. The factor $a$ is a tuning parameter which determines the relative weight of the mass and the geometry of the reconstructed decay in the minimization, with a value of 40 used in the present study. The chosen value provided the best performance in a coarse parameter scan, however a rigorous optimization of this parameter has not yet been performed. The position which gives the lowest $\chi^2$ is taken as the reconstructed position of the neutral pion. The combination of mass and photon direction breaks the degeneracy of possible pion positions present when using mass alone, and thus enables stand-alone neutral pion reconstruction in the calorimeter. Here, only one particle is simulated per event, yielding exactly two photon clusters. In more realistic environments, the absolute value of the minimum $\chi^2$ can also be used to establish the assignment of photons to neutral pion candidates, and to give a likelihood for the candidate to be a true $\pi^0$. With precise timing information in combination with the determined decay position the association of photons to $\pi^0$ candidates can also be improved.

\begin{figure}
\centering
\includegraphics[width = 0.45\textwidth]{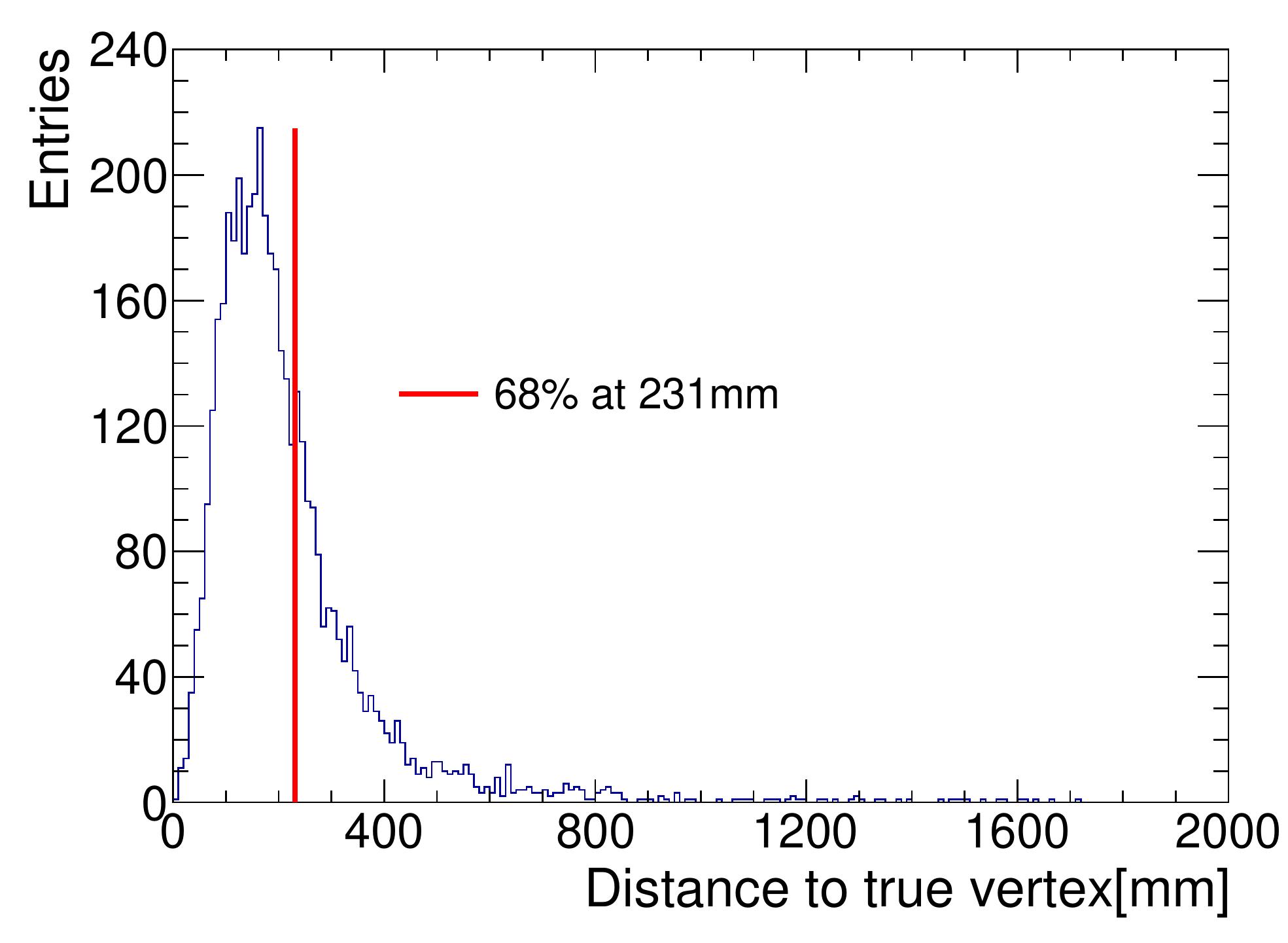} \hspace{5mm}
\includegraphics[width = 0.44\textwidth]{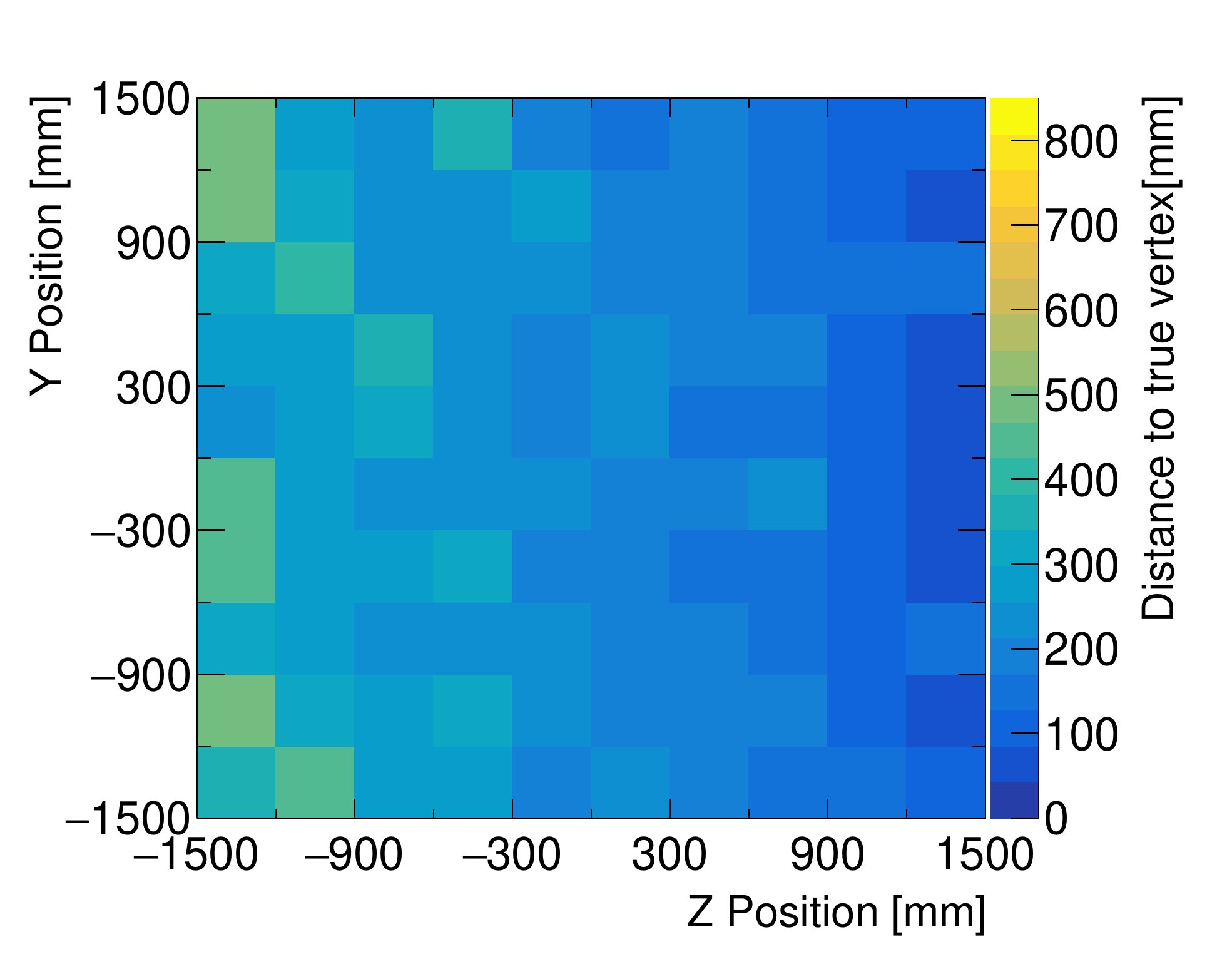}
\caption{Left: Three dimensional distance of reconstructed and true $\pi^0$ position for neutral pions with 400\ MeV kinetic energy in z direction, decaying in the center of the assumed detector volume, 1.5\ m away from the calorimeter. The precision of the position reconstruction, given by the 68\%ile of the distribution, is indicated by the red vertical line. Right: Distribution of precision of the position reconstruction for neutral pions with 400\ MeV kinetic energy in z direction over the full detector volume. \label{fig:Pi0Reco}}
\end{figure}

\autoref{fig:Pi0Reco} shows the resolution of the decay position of neutral pions with a kinetic energy of 400\ MeV traveling in the z direction. The distribution of the distance between true and reconstructed position is shown in detail for pions decaying in the center of the detector, 1.5\ m away from the front face of the downstream calorimeter module, and the resolution, given by the 68\%ile of the distribution, over the full detector volume. It is apparent that the location of neutral pions can be determined with a precision of 10 to 30\ cm, depending on the distance from the downstream calorimeter. The minimization procedure improves the accuracy by more than a factor of two compared to a purely geometric reconstruction. For lower pion energies the reconstruction depends more strongly on the distance from the downstream calorimeter, and is less accurate. For energies above 400\ MeV no significant additional improvement is observed. 

\subsection{Neutrons}

The calorimeter is also sensitive to neutrons. The probability for an interaction which deposits at least 0.5\ MeV in one detector cell varies between 75\% and 95\% for neutrons with kinetic energies from 10\ MeV to 550\ MeV, with the highest interaction probability observed for energies around 40\ MeV. Since such single hit events are in practice very likely not useable, a requirement of at least 5 hits with a total visible energy of at least 5\ MeV is introduced to quantify the neutron sensitivity of the ECAL concept. This substantially reduces the efficiency at neutron energies below 100\ MeV.  It should be noted that no constraints are imposed on the geometrical relation of these hits, so it does not imply that a clear cluster of energy is identifiable. For a more realistic analysis of the neutron reconstruction capability more sophisticated algorithms will be needed. 

\begin{figure}
\centering
\includegraphics[width = 0.49\textwidth]{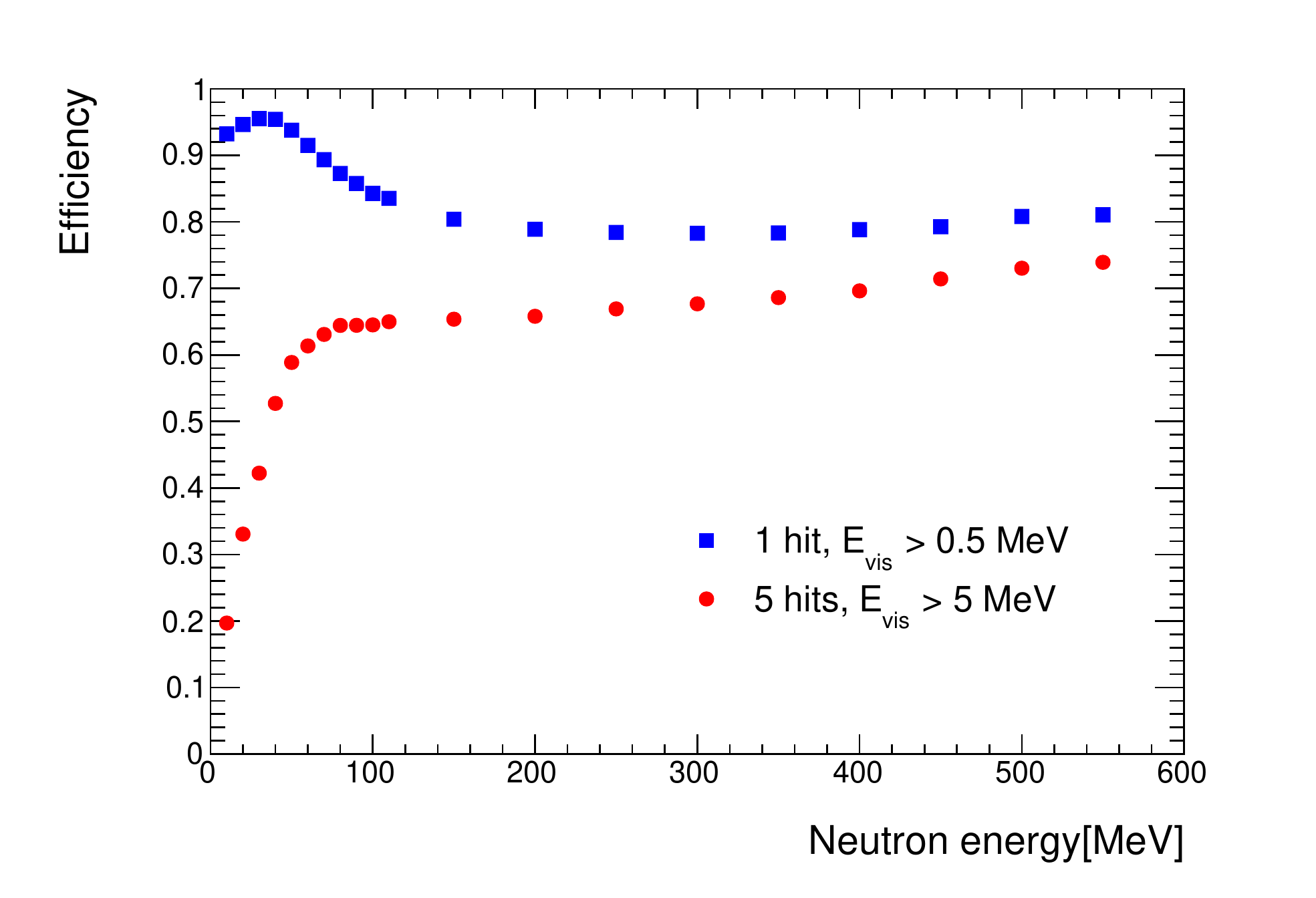} \hfill
\includegraphics[width = 0.44\textwidth]{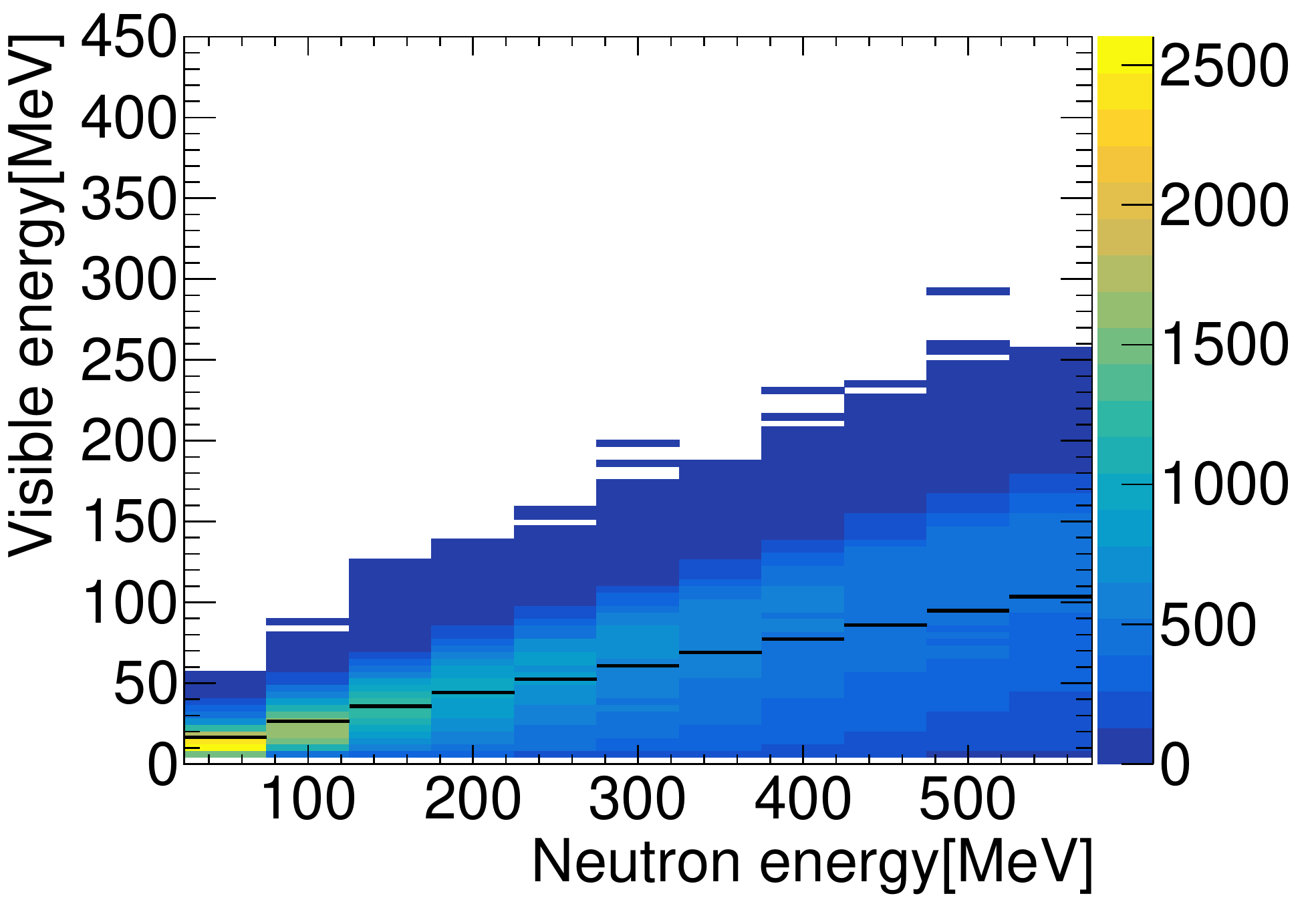}
\caption{Left: Fraction of neutrons detected by the calorimeter for a minimal threshold of one cell above 0.5\ MeV visible energy, and a stricter requirement of five hits and at least 5\ MeV visible energy. Right: Distribution of the visible energy as a function of neutron energy. \label{fig:Neutrons}}
\end{figure}

\autoref{fig:Neutrons} shows the efficiency for neutron detection both for the minimum and the stricter requirements outlined above, and the distribution of the visible energy as a function of neutron energy. It is apparent that there is significant potential for the detection of neutrons by the ECAL, but the energy measurement will not be very accurate. With nanosecond timing, time of flight measurements could be used to improve the energy estimate when the precise time of the relevant neutrino interaction, given for example by other particles entering the calorimeter, is known.

\section{Towards a more realistic concept}
\label{sec:realism}

The present study illustrates the potential of a highly granular electromagnetic calorimeter with shower imaging capabilities for the DUNE near detector. In the studied default configuration, with 80 layers and a cell size of $20 \times 20\ \mathrm{mm}^2$, such a calorimeter would have 200\,000 channels per square meter, resulting in several tens of millions of channels for a full near detector system. This leads to unrealistically high costs and complexity. The increase of the granularity to $30 \times 30\ \mathrm{mm}^2$ would result in a reduction of the channel count by more than a factor of two with a moderate penalty of the angular resolution of the system. The studies discussed here have demonstrated that the granularity is primarily important in the first layers of the detector, enabling the transition to crossed scintillator strips rather than individual scintillator tiles in later layers. This would result in significant reductions of the channel count. Presently ongoing studies also investigate the impact of thicker active and passive layers, such as 10\ mm thick scintillator tiles combined with 4 or 5\ mm thick copper plates, which would reduce the number of layers, and thus the number of channels, by a factor of two. These studies are accompanied by experimental studies to establish the viability of the direct coupling of SiPMs also to such thicker scintillator tiles. In addition, the full thickness of the detector may not be needed all around the tracking volume, as demonstrated by the design of the T2K ND280 ECAL \cite{Abe:2011ks, Allan:2013ofa}. Together, these modifications would bring the channel count of the calorimeter system down to an order of one to a few million channels, in reach of present technology when making use of automated assembly techniques for the highly granular layers. 

On the simulation side, additional realism, in particular concerning material for the SiPM readout, the inclusion of non-uniformities and material inhomogeneities as well a a more sophisticated implementation of digitization effects and reconstruction algorithms, is needed to further advance the detector optimizations. These efforts are now also beginning. 

\section{Conclusions}

First simulation studies have shown that a highly granular SiPM / plastic scintillator based sampling ECAL provides significant potential for the reconstruction of neutrino events in the near detector of the DUNE experiment, in particular by its capability to determine the location of neutral pions in the tracker volume of the detector. It also provides significant neutron detection efficiency. Both are particularly relevant in the scenario with a high pressure gaseous TPC as main tracking detector. Further studies with increased realism, both in terms of the detector design and in the degree of sophistication of the modeling in simulations, are ongoing to fully establish this detector concept as a viable option for the DUNE near detector.

\section*{References}
\bibliography{DUNE_ECAL}

\end{document}